\documentclass[12pt]{article}
\usepackage[sc]{mathpazo}
\usepackage{graphicx, amsmath}
\usepackage{verbatim}
\usepackage{hyperref}
\usepackage{animate}
\usepackage{amssymb}
\usepackage{titling}  
\usepackage{url}
\usepackage[a4paper]{geometry}
\usepackage{algorithmic}
\usepackage{arydshln}
\usepackage{float}
\usepackage{natbib}
\usepackage{color}
\usepackage{fancyvrb}
\usepackage{setspace}
\usepackage{wasysym}
\usepackage[margin=10pt, font=scriptsize, labelfont=bf]{caption}

\numberwithin{equation}{section}
\usepackage{anysize}
\marginsize{25mm}{25mm}{25mm}{25mm}

\date{}
\begin{document}
\title{\textbf{STAD Research Report 03\_2015} \\
\vspace{12pt}
Adjusted Concordance Index, an extension of the Adjusted Rand index to fuzzy partitions}
\author{
Sonia Amodio$^{a}$, Antonio d'Ambrosio$^{a}$, Carmela Iorio$^{a}$, Roberta Siciliano$^{b}$, \\
\\
$^{a}$ Department of Economics and Statistics,\\
University of Naples Federico II,\\
Naples, Italy.\\
\href{sonia.amodio}{\{sonia.amodio, \ antdambr, \ carmela.iorio \} @unina.it}
\\
$^{b}$ Department of Industrial Engineering,\\
University of Naples Federico II,\\
Naples, Italy.\\
\href{roberta@unina.it}{ roberta@unina.it} \\
}
 \maketitle

\begin{abstract}
In comparing clustering partitions, Rand index (RI) and Adjusted Rand index (ARI) are commonly used for measuring the agreement between the partitions. Both these external validation indexes aim to analyze how close is a cluster to a reference (or to prior knowledge about the data) by counting corrected classified pairs of elements.
When the aim is to evaluate the solution of a fuzzy clustering algorithm, the computation of these measures require converting the soft partitions into hard ones. It is known that different fuzzy partitions describing very different structures in the data can lead to the same crisp partition and consequently to the same values of these measures.\\
\noindent We compare the existing approaches to evaluate the external validation criteria in fuzzy clustering and we propose an extension of the ARI for fuzzy partitions based on the \textit{normalized degree of concordance}. Through use of real and simulated data, we analyze and evaluate the performance of our proposal.

\end{abstract}

\textbf{keywords}: Clustering, cluster validity, fuzzy partitions, external evaluation measures, Rand index, Adjusted Rand index

\section{Introduction}
\label{intro}
\noindent Cluster analysis, broadly speaking, can be defined as an unsupervised method to partition a set of objects, $\mathbf{X} = \{x_{i,j}\}_{(n \times p)}$, into a finite set of clusters, $\mathbf{C} = \{\mathbf{C}_1, ..., \mathbf{C}_K\}$, according to the similarities among these objects \citep{anderberg,berkhin,bock,duran,fasulo,hartigan,kaufman,jain88,jain99,mirkin,spath}, in such a way that objects in the same cluster are as similar as possible and objects in different clusters are as dissimilar as possible.
An important distinction can be made between hard and soft clustering algorithms.
Hard clustering methods consider disjoint partitions. In other words, an object belongs or does not belong to a cluster. More formally, given a data set $\mathbf{X}$, the clustering structure can be presented as a set of non empty subsets $\{\mathbf{C}_1, \cdots, \mathbf{C}_k,  \cdots, \mathbf{C}_K\}$ such that:

\begin{equation}
 \begin{split}
        X = \bigcup_{k=1}^{K} \mathbf{C}_k, \\
        \mathbf{C}_k \bigcap \mathbf{C}_{k'} = \varnothing, \quad \mbox{for} \quad k\neq k'.
 \end{split}
\end{equation}

\noindent When the data presents no sharp boundaries between clusters, fuzzy clustering algorithms should be preferred. These methods determine for each object a degree of membership to belong to every cluster \citep{ruspini,bezdek,hoppner}. In this way, objects that are on the boundary between different clusters are not forced to belong to one specific cluster, but they present a different degree of membership for each cluster.  More formally, these methods partition the elements in $\mathbf{X}$ in $K$ fuzzy overlapping clusters, with respect to some defined criterion, and they returns both a set of cluster centers and a partition matrix of the following form
\begin{equation}
\mathbf{W} = \{w_{i,k} \}_{(n \times K)} \in [0,1]; \quad \sum_{k=1}^K w_{i,k} = 1 \quad \forall i \in \{1, \cdots, n\},
\end{equation}

\noindent in which $w_{i,k}$ represents the degree to which the element $\mathbold{x}_i$ belongs to the cluster $\mathbold{C}_k$. \

\noindent Cluster validation involves both internal and external validation criteria. The internal validation criteria are based only on the observation on the clustered data while the external ones are defined on some information that is not used in the clustering production, i.e.\ a golden standard cluster structure known a priori. Several external validation criteria have been proposed in the literature. These scalar indexes assess the goodness of the partition obtained by the clustering procedure on the base of previous knowledge about the data. Among these measures we can mention the Rand index \citep{rand}, the Fowlkes and Mallows $\mathcal{F}$-measure \citep{fowlkes}, the Jaccard index \citep{downton}, the Mirkin metric \citep{mirkin} and the Dice coefficient \citep{dice}. \\
\noindent The remainder of the paper is organized as follows: in Section \ref{rand} the Rand index and the adjusted version by \citet{hubert} are presented. In Section \ref{fuzzyrand} some approaches to extend the Rand index and the Adjusted Rand index to fuzzy partitions are presented. Section \ref{ourmeth} is devoted to introduce and explain the Adjusted Concordance index. Section \ref{sim} is dedicated to the analysis of the performance of the proposed index. Concluding remarks close the paper in Section \ref{final}.

\section{Rand index, Adjusted Rand index and related measures}
\label{rand}
\noindent The Rand index is an \textit{external evaluation measure} developed by \citet{rand} to compare the clustering partitions on a set of data. \\
Let $\mathbf{X} = \{x_{ij}\}_{(n \times p)}$ be the data matrix, where $n$ is the number of objects and $p$ the number of variables. A partition of the $n$ objects in $K$  subsets or groups, $\mathbf{P} = \{\mathbf{P}_1,, ..., \mathbf{P}_K\}$, can be formed in such a way that the union of all the subsets is equal to the entire data set and the intersection of any two subsets is the empty set. \\
It is possible to say that two elements of $\mathbf{X}$, i.e.\ $(\mathbf{x}, \mathbf{x'})$ are paired in $\mathbf{P}$ if they belong to the same cluster.
Let $\mathbf{P}$ and $\mathbf{Q}$ be two partitions of the objects set $\mathbf{X}$. The Rand index is calculated as:
\begin{equation}
RI = \frac{a + d} {a + b + c + d} = \frac{a+d}{\binom {n} {2}},
\end{equation}
where
\begin{itemize}
\item $a$ is the number of pairs $(\mathbf{x}, \mathbf{x'}) \in \mathbf{X}$ that are paired in $\mathbf{P}$ and in $\mathbf{Q}$;
\item $b$ is the number of pairs $(\mathbf{x}, \mathbf{x'}) \in \mathbf{X}$ that are paired in $\mathbf{P}$ but not paired in  $\mathbf{Q}$;
\item $c$ is the number of pairs $(\mathbf{x}, \mathbf{x'}) \in \mathbf{X}$ that are not paired in $\mathbf{P}$ but paired in $\mathbf{Q}$;
\item $d$ is the number of pairs $(\mathbf{x}, \mathbf{x'}) \in \mathbf{X}$ that are neither paired in $\mathbf{P}$ nor in $\mathbf{Q}$.
\end{itemize}
This index varies in $[0, 1]$ with $0$ indicating that the two partitions do not agree on any pair of elements and $1$ indicating that the two partitions are exactly the same. Unfortunately the Rand statistic approaches its upper limit as the number of clusters increases.\

\noindent There are some other known problems with the Rand index \citep{meilua}:
\begin{itemize}
\item The expected value of the Rand index for two random partitions does not take a constant value;
\item It presents high variability and, as proved by \citet{fowlkes}, it concentrates in a small interval close to 1;
\item It is extremely sensitive to the number of groups considered in each partition (as proved by \citet{morey}), their size and also to the overall number of observations considered.
\end{itemize}
To overcome these problems \citet{hubert} proposed a corrected version of the Rand index assuming the generalized hypergeometrical distribution as model of randomness (i.e.\ $\mathbf{P}$ and $\mathbf{Q}$ are picked at random with a fixed number of partitions and a fixed number of elements in each). In other words, this corrected version is equal to the normalized difference of the Rand index and its expected value under the null hypothesis:
\begin{equation}
ARI = \frac{Index - Expected \; Index}{Maximum \; Index - Expected \; Index}.
\end{equation}
For further details we refer to \cite{hubert}.
More formally the Hubert-Arabie's formulation of the adjusted Rand index is:
\begin{equation}
\label{ari}
ARI_{HA} = \frac{2(ad - bc)}{b^2 + c^2 + 2ad + (a+d)(c+b)}.
\end{equation}
This index has an upper bound of 1 and takes the value 0 when the Rand index is equal to its expected value (under the generalized hypergeometric distribution assumption for randomness). Negative values are possible but not interesting since they indicate less agreement than expected by chance.\\
As regarding the other related comparison measures, all of them can also be expressed in terms of the four cardinalities $a, b, c$ and $d$.
The Jaccard index, also known as Tanimoto coefficient, is equal to $\mathcal{J} = \frac{a}{a+b+c}$, the Fowlkes-Mallow $\mathcal{F}$-index is equal to $\mathcal{F} = \frac{a}{\sqrt{(a+b)(a+c)}}$, the Mirkin metric can be written as $\mathcal{M} = 2(b+c)$ and the Dice coefficient is equal to $\frac{2a}{2a+b+c}$. For a more extensive description of these related indexes we refer to \cite{wagner}.

\section{Extensions of the Rand index to fuzzy partitions}
\label{fuzzyrand}

\noindent The problem of evaluating the solution of a fuzzy clustering algorithm with the Rand index is that it requires converting the soft partition into a hard one, in this way, losing information. As shown in \cite{campello}, different fuzzy partitions describing different structures in the data may lead to the same crisp partition and then in the same Rand index value. For this loss of information the Rand index is not able to discriminate between overlapping and non-overlapping clusters. Therefore it is not appropriate for fuzzy clustering assessment.\\
\citet{campello} proposed a fuzzy extension of the Rand index and related indexes by defining a set-theoretic form to calculate the four cardinalities. His main goal was to compare a fuzzy partition with a non-fuzzy one, but as he notes himself, his measure can also used to compare two fuzzy partitions. Unfortunately this measure fails in satisfying reflexivity (i.e. the extension of the Rand index calculated between two identical partitions is less than 1), and thus it cannot be considered a proper metric.
\citet{frigui} proposed a similar measure, which can be considered a special case of Campello's one, that also cannot be considered a proper metric since it also fails in satisfying reflexivity.
\citet{brouwer} proposed an alternative extension of the Rand index and related measures, based on the cosine correlation as measure of bonding (or similarity) between two items with fuzzy membership vectors. Also this measure unfortunately violets the reflexivity condition.

\citet{anderson} proposed a fuzzy generalization of the Rand index and other measures between soft partitions (i.e.\ fuzzy and possibilistic partitions) based on matrix operations that presents a clear advantage in terms of efficiency since it does not consider all pairs of objects involved in the calculation of the four cardinalities necessary to calculate these indexes.  Unfortunately, also in this case, as the authors stated, their generalization is a similarity measure that cannot be interpreted as a metric. For a more extensive discussion on these aforementioned proposals, we refer to \cite{huller}.\\
\citet{huller} proposed a generalization of the Rand index and of the related measures, namely the Jaccard measure and the Dice coefficient. \\
Let $\mathbf{P} = \{\mathbf{P}_1, ..., \mathbf{P}_K\}$ be a fuzzy partition of the data matrix $\mathbf{X}$. Each item $\mathbf{x} \in \mathbf{X}$ is then characterized by its membership vector:
\begin{equation}
\mathbf{P}(x) = (\mathbf{P}_1(\mathbf{x}), \mathbf{P}_2(\mathbf{x}), ..., \mathbf{P}_K(\mathbf{x})) \vspace{3mm} \in [0,1]^{K},
\end{equation}
where $\mathbf{P}_i(\mathbf{x})$ is the membership degree of $\mathbf{x}$ in the $i$-th cluster $\mathbf{P}_i$.
Given any pair $(\mathbf{x}, \mathbf{x}') \in \mathbf{X}$, they defined a fuzzy equivalence relation on $\mathbf{X}$ in terms of similarity measure as:
\begin{equation}
E_{\mathbf{P}} = 1 -  \|\mathbf{P}(\mathbf{x}) - \mathbf{P}(\mathbf{x}')\|,
\end{equation}
where  $\|\cdot\|$ is the normalized $L_1$-norm, which constitutes a proper metric on $[0,1]^{K}$ and yields value in $[0,1]$. $E_\mathbf{P}$ is equal to 1 if and only if $x$ and $x'$ have the same membership pattern and is equal to 0 otherwise.\\
Given 2 fuzzy partition, $\mathbf{P}$ and $\mathbf{Q}$, the basic idea underneath the fuzzy extension of the Rand index is to generalize the concept of \textit{concordance} in the following way.
Considering a pair $(\mathbf{x}, \mathbf{x}')$ as being concordant as $\mathbf{P}$ and $\mathbf{Q}$ agree on its degree of equivalence, \cite{huller} define the \textit{degree of concordance} as:
\begin{equation}
conc(\mathbf{x}, \mathbf{x}') = 1 - \|E_{\mathbf{P}}(\mathbf{x}, \mathbf{x}') - E_{\mathbf{Q}}(\mathbf{x}, \mathbf{x}')\| \vspace{3mm} \in [0,1],
\end{equation}
and the \textit{degree of discordance} as:
\begin{equation}
disc(\mathbf{x}, \mathbf{x}') = \|E_{\mathbf{P}}(\mathbf{x}, \mathbf{x}') - E_{\mathbf{Q}}(\mathbf{x}, \mathbf{x}')\|
\end{equation}
The distance measure is then defined by the normalized sum of concordant pairs:
\begin{equation}
d(\mathbf{P},\mathbf{Q}) = \frac{\sum_{(\mathbf{x}, \mathbf{x}') \in \mathbf{X}}{\|E_{\mathbf{P}}(\mathbf{x}, \mathbf{x}') - E_{\mathbf{Q}}(\mathbf{x}, \mathbf{x}')\|}}{n(n-1)/2}.
\end{equation}

\noindent The direct generalization of the Rand index corresponds to the normalized degree of concordance (NDC) and it is equal to:
\begin{equation}
R_E(\mathbf{P},\mathbf{Q}) = 1 - d(\mathbf{P},\mathbf{Q}),
\end{equation}
and it reduces to the original Rand index when partitions $\mathbf{P}$ and $\mathbf{Q}$ are non-fuzzy.\\
\noindent This distance is a pseudo-metric, since it always satisfies the conditions of non-negativity, reflexivity, symmetry, triangle inequality and it is a metric when we consider particular assumptions (which can be summarized in considering Ruspini's partitions, the existence of a prototypical element for each cluster and the equivalent relation on $\mathbf{X}$: $E_\mathbf{P}(\mathbf{x}, \mathbf{x}') = 1 - \| \mathbf{P}(\mathbf{x} - \mathbf{P}(\mathbf{x}')$) because it also satisfies the separation condition). \\
Since we are interested in comparing fuzzy partitions and the adjusted Rand index proposed by Hubert and Arabie is still the most popular measure used for clusterings comparison, we propose an extension of this index to fuzzy partitions, namely the Adjusted Concordance Index, based on the fuzzy variant of the Rand index proposed by \cite{huller}.
These authors indeed proposed the extension of a large number of related comparison measures, which can be expressed in terms of the cardinals $a, b, c$ and $d$, through the formalization of these cardinals in fuzzy logic concordance terms.\\
These cardinals can be expressed as follows:
\begin{itemize}
\item $a$-concordance: objects $x$ and $x'$ are concordant because their degree of equivalence in $\mathbf{P}$ and in $\mathbf{Q}$ is similar and their degree of equivalence in $\mathbf{P}$ is high and their degree of equivalence in $\mathbf{Q}$ is high
\begin{equation*}
a = \top(1 - |E_\mathbf{P}(x,x')-E_\mathbf{Q}(x,x')|, \top(E_\mathbf{P}(x,x'),E_\mathbf{Q}(x,x'));
\end{equation*}
\item $d$-concordance: negation of a-concordance (objects $x$ and $x'$ are concordant but either the degree of equivalence in $\mathbf{P}$ is not high or the degree of equivalence in $\mathbf{Q}$ is not high)
\begin{equation*}
d = \top(1 - |E_\mathbf{P}(x,x')-E_\mathbf{Q}(x,x')|,\bot (1 - E_\mathbf{P}(x,x'), 1 - E_\mathbf{Q}(x,x'));
\end{equation*}
\item $b$-discordance: the degree of equivalence of $x$ and $x'$ in $\mathbf{P}$ is larger than that in $\mathbf{Q}$
\begin{equation*}
b = \max (E_\mathbf{P}(x,x')-E_\mathbf{Q}(x,x'),0);
\end{equation*}
\item $c$-discordance: the degree of equivalence of $x$ and $x'$ in $\mathbf{P}$ is smaller than that in $\mathbf{Q}$
\begin{equation*}
c = \max (E_\mathbf{Q}(x,x')-E_\mathbf{P}(x,x'),0);
\end{equation*}
\end{itemize}
where $\top$ is the triangular product norm and $\bot$ is the associated triangular conorm (algebraic sum) \citep{klement}.
The cardinals just mentioned can be also expressed as:
\begin{subequations}
    \begin{align}
a &= (1-|E_\mathbf{P}(x,x') - E_\mathbf{Q}(x,x')|) \cdot E_\mathbf{P}(x,x') \cdot E_\mathbf{Q}(x,x') \notag \\
d &= (1 - |E_\mathbf{P}(x,x') - E_\mathbf{Q}(x,x')|) \cdot (1 - E_\mathbf{P}(x,x') \cdot E_\mathbf{Q}(x,x'))\notag \\
b &= max(E_\mathbf{P}(x,x') - E_\mathbf{Q}(x,x'), 0) \notag \\
c &= max(E_\mathbf{Q}(x,x') - E_\mathbf{P}(x,x'), 0)  \notag \\
    \end{align}
\end{subequations}


\section{The adjusted concordance index}
\label{ourmeth}
\noindent \cite{huller} did not propose explicitly an extension of the adjusted Rand index. Our idea of an extension of the normalized degree of concordance (NDC) was born when we noticed that the other proposed extensions of the Rand index to fuzzy partitions were based upon the generalization of the four cardinalities presented in a standard contingency table to compare 2 partitions.
\citeauthor*{huller}'s proposal of a fuzzy version of the Rand index instead is based on the fuzzy equivalence relation and this allows us to rewrite every partition as a similarity matrix based on the normalized city block distance.
For example, if we consider the following two crisp partitions \\
\begin{center}
$
\mathbf{P} = \begin{bmatrix}
       1 & 0 \\
       1 & 0 \\
       0 & 1 \\
			 1 & 0 \\
     \end{bmatrix}
$ and  $
\mathbf{Q} = \begin{bmatrix}
       1 & 0 \\
       0 & 1 \\
       0 & 1 \\
			 1 & 0 \\
     \end{bmatrix}
		$, \\
	\end{center}
		\noindent we obtain that $E_\mathbf{P}$ and $E_\mathbf{Q}$ are equal to \\
\begin{center}
$
\mathbf{E_\mathbf{P}} = \begin{bmatrix}
       1 & 1 & 0 & 1 \\
       1 & 1 & 0 & 1 \\
       0 & 0 & 1 & 0 \\
			 1 & 1 & 0 & 1 \\
     \end{bmatrix}
		$ $\;$ , $\;$
$
\mathbf{E_\mathbf{Q}} = \begin{bmatrix}
       1 & 0 & 0 & 1 \\
       0 & 1 & 1 & 0 \\
       0 & 1 & 1 & 0 \\
			 1 & 0 & 0 & 1 \\
     \end{bmatrix}
		$.
\end{center}
\noindent This enables us to calculate the four cardinalities, $a, b, c$ and $d$, by considering pairs of objects that are paired in both partitions, pairs of objects that are not paired in both partitions, and pairs of objects that are paired in a partition but not in the other and vice versa, obtaining, in this simple example: $a = 1$, $b = 2$, $c = 1$ and $d = 2$.
Using the standard formulation of the contingency table to compare two partitions, it is possible to obtain exactly the same values for the four cardinalities \citep{hubert, meilua} Then we can see that the Rand index and the adjusted Rand index for this toy example are respectively equal to $RI = 0.5$ and $ARI = 0$.

\noindent Following the same line of reasoning, when we consider fuzzy partitions  the elements in the matrices $E_\mathbf{P}$ and $E_\mathbf{Q}$ are, of course, real numbers between $0$ and $1$ and they represent respectively similarities between pairs of objects in the same partition. Consequently, the normalized degree of concordance can be seen as the similarity measure between two similarity matrices. For instance, considering two random fuzzy partitions of a set of $n=4$ objects\\

\begin{center}
$\mathbf{P}' = \begin{bmatrix}
       0.29 & 0.71 \\
       0.79 & 0.21 \\
       0.41 & 0.59 \\
			 0.88 & 0.12 \\
     \end{bmatrix}$
and $
\mathbf{Q}' = \begin{bmatrix}
       0.94 & 0.06 \\
       0.05 & 0.95 \\
       0.53 & 0.47 \\
			 0.89 & 0.11 \\
     \end{bmatrix}$, \\
\end{center}		
\noindent we obtain that $E_{\mathbf{P}'}$ and $E_{\mathbf{Q}'}$ corresponds to \\

\begin{center}
$E_{\mathbf{P}'} = \begin{bmatrix}
  1.00 & 0.50 & 0.88 & 0.41 \\
  0.50 & 1.00 & 0.62 & 0.91 \\
  0.88 & 0.62 & 1.00 & 0.53 \\
  0.41 & 0.91 & 0.53 & 1.00 \\
	\end{bmatrix}$
and $E_{\mathbf{Q}'} = \begin{bmatrix}
	1.00 & 0.11 & 0.59 & 0.95 \\
  0.11 & 1.00 & 0.52 & 0.16 \\
  0.59 & 0.52 & 1.00 & 0.64 \\
  0.95 & 0.16 & 0.64 & 1.00 \\
	\end{bmatrix}$,
	\end{center}
\noindent resulting in a $NDC = 0.6367$. \\
\noindent It can be noted that $E_\mathbf{P}$ and $E_\mathbf{Q}$ are symmetric matrices with a number of unique similarities equal to $m = \frac{n(n-1)}{2}$.\\
\noindent Unfortunately it is not possible to adjust the NDC by considering an adjustment of the four cardinalities. The formulation of the Adjusted Rand index by \cite{hubert} in terms of the four cardinalities, reported in Equation \ref{ari}, was derived as a simplification of
\begin{equation}
ARI_{HA} = \frac{\sum_{i,j} \binom {n_{ij}}{2} - \sum_i\binom{n_{i+}}{2} \sum_j \binom{n_{+j}}{2} / \binom{n}{2}}
{0.5 [ \sum_i\binom{n_{i+}}{2} + \sum_j\binom{n_{+j}}{2} ] - \sum_i\binom{n_{i+}}{2} \sum_j \binom{n_{+j}}{2} / \binom{n}{2}},
\end{equation}
where $n$ is the number of objects, $n_{i+}$ and $n_{+j}$ are respectively the rows and the columns marginals of the contingency table obtained by crossing two crisp partition vectors. When dealing with fuzzy partitions it is not straightforward to obtain contingency tables. \cite{anderson}), following the same approach of \cite{hubert}, obtained a fuzzy generalization of the contingency tables, but the drawback of this generalization is that neither the marginals nor the elements of the tables are integers and then some of the cardinalities can be negative. This fact makes not straightforward the use of the binomial coefficients. \\
\noindent Therefore, the key idea is to use the NDC and normalize the difference between the NDC and its expected value. We estimate the expected value of the NDC by considering the average value of the index after permuting the elements of each upper triangular similarity matrices, given the partitions and a certain number of groups. When the number of pairwise similarities ($m$) is small, the estimate of the expected value is based on considering all possible permutations ($m!$), while, when the number of the pairwise similarities is large, we estimate the expected value by taking in account $h$ randomly selected permutations on the total $m!$ permutations. \\
For the toy example, we have that the expected value of the NDC, considering all possible permutations of the upper triangular similarity matrices ($m! = 720$), corresponds to 0.6972. Hence, the ACI is equal to

\begin{equation}
ACI = \frac{NDC - expected \; NDC}{1 - expected \; NDC} = \frac{0.6367 - 0.6972}{1 - 0.6972} = -0.200.
\end{equation}

\noindent Similarly to \citeauthor*{hubert}'s ARI, negative values of the ACI are possible but not interesting since they indicate less agreement than expected by chance and then the index can be set equal to zero. \\
\noindent It is worth stressing that we can correct the NDC according to the proposed approach because it is the only extension of the Rand index to fuzzy partitions which fulfill the reflexivity property that always guarantees that its maximum value is equal to one.\\
\noindent It is worth mentioning that we made an experiment to evaluate the bias in our estimate of the expected value of the NDC. In this experiment we generated 1000 data sets with a random sample size $n$ between 100 and 1200, a random number of clusters $C$, randomly chosen between 2 and 10, a random number of dimensions between 2 and 10, and a covariance matrix $\Sigma = I\alpha$ with $\alpha$ between 0.1 and 3. We stored for each data set the composition of the clusters and we performed a cluster analysis using the K-means algorithm. For each solution we calculated the \citeauthor{hubert}'s adjusted Rand index and the ACI. The average difference between these two measures resulted to be $-4.5602 \times 10^{-07}$.


\section{Experimental evaluation through simulated and real data set analyses}
\label{sim}
		
\subsection{Comparing fuzzy and crisp partitions}
\label{fuzzycrisp}

For the first simulation study, we generated data with $\mathbf{C}=2,3,4$ cluster centers by incrementally merging four different bi-variate normal distributions with mean vectors

\begin{center}
\begin{tabular}{l}
$\boldsymbol{\mu}_1 = [-2,-2];$ \\
$\boldsymbol{\mu}_2 = [2,2];$ \\
$\boldsymbol{\mu}_3 = [0,0];$ \\
$\boldsymbol{\mu}_4 = [-2,2],$ \\
\end{tabular}
\end{center}

\noindent and three different levels of variability described by the following covariance matrices: $\boldsymbol{\Sigma}_1=\textrm{I}\times 0.01$, $\boldsymbol{\Sigma}_2=\textrm{I}\times 0.25$, $\boldsymbol{\Sigma}_3=\textrm{I}\times 1$. The structure of the simulated data sets is presented in Table \ref{tab3}. The sample size was set equal to 100.

\begin{table}[ht]
\centering
\caption{First simulation study 1: data sets structure}
\label{tab3}
\begin{tabular}{|l|l|}
\hline
\multicolumn{1}{|c|}{Data set number} & \multicolumn{1}{c|}{Centers} \\
\hline
\multicolumn{1}{|c|}{1} & $\boldsymbol{\mu}_1, \textrm{ }\boldsymbol{\mu}_2$ \\
\hline
\multicolumn{1}{|c|}{2} & Add $\boldsymbol{\mu}_3$ to data set 1\\
\hline
\multicolumn{1}{|c|}{3} & Add $\boldsymbol{\mu}_4$ to data set 2 \\
\hline
\end{tabular}
\end{table}

\noindent We then generated three data sets with $\mathbf{C}=2,3,4$ by sampling at turns from the same bi-variate normal distribution with mean vector equal to $[0,0]$ and covariance equal to $\textrm{I}\times 0.8$. For each data set, we stored the crisp membership matrix and we ran the fuzzy \textit{C}-mean algorithm for each of them by setting $\mathbf{C}=2,3,4$.
We computed the normalized degree of concordance (NDC), the extension of the Rand index proposed by \cite{brouwer}, by \cite{campello} and by \cite{anderson}.


    \begin{table}[ht]
    \centering
		\caption{Extensions of the Rand index. Comparison between H\"ullermeier el al., Brouwer, Campello, Anderson et al.}
		\label{table:frand}
    \begin{tabular}{|c|c|c|c|c|}
    \hline
     & NDC & Brouwer & Campello & Anderson \\
    \hline
    2 Centers, $\Sigma_1$ & 0.9992 & 0.9995 & 0.9992 & 0.9989 \\
    \hline
    2 Centers, $\Sigma_2$ & 0.9777 & 0.9846 & 0.9779 & 0.9709 \\
    \hline
    2 Centers, $\Sigma_3$ & 0.9097 & 0.9280 & 0.9101 & 0.8857 \\
    \hline
    3 Centers, $\Sigma_1$ & 0.9961 & 0.9977 & 0.9970 & 0.9955 \\
    \hline
    3 Centers, $\Sigma_2$ & 0.9487 & 0.9630 & 0.9511 & 0.9417 \\
    \hline
    3 Centers, $\Sigma_3$ & 0.8545 & 0.8617 & 0.8486 & 0.8393 \\
    \hline
    4 Centers, $\Sigma_1$ & 0.9947 & 0.9972 & 0.9968 & 0.9944 \\
    \hline
    4 Centers, $\Sigma_2$ & 0.8820 & 0.9125 & 0.9154 & 0.8888 \\
    \hline
    4 Centers, $\Sigma_3$ & 0.7187 & 0.7350 & 0.7639 & 0.7598 \\
    \hline
    Random 2 Centers & 0.4963 & 0.4954 & 0.4979 & 0.4953 \\
    \hline
    Random 3 Centers & 0.4982 & 0.4720 & 0.5318 & 0.5527 \\
    \hline
    Random 4 Centers & 0.5231 & 0.4910 & 0.5721 & 0.6207 \\
    \hline
    \end{tabular}
    \end{table}
		
\noindent As can be noted from Table \ref{table:frand}, the behavior of the extensions of the Rand index in all simulated data sets is quite similar when these indexes are used to compare a fuzzy partition with the known crisp partition. We also computed the ACI and the extensions of the Adjusted Rand index proposed by \cite{brouwer, campello} and \cite{anderson}. Unsurprisingly, as can be noted from Table \ref{table:fadjst}, a conclusion similar to the previous one can be reached.

    \begin{table}[ht]
    \centering
    \caption{Fuzzy extensions of adjusted Rand index. Comparison between ACI, Brouwer, Campello, and Anderson et al.}
    \begin{tabular}{|c|c|c|c|c|}
    \hline
     & ACI & Brouwer & Campello & Anderson \\
    \hline
    2 Centers, $\Sigma_1$ & 0.9984 & 0.9989 & 0.9984 & 0.9978 \\
    \hline
    2 Centers, $\Sigma_2$ & 0.9555 & 0.9693 & 0.9557 & 0.9419 \\
    \hline
    2 Centers, $\Sigma_3$ & 0.8196 & 0.8561 & 0.8202 & 0.7714 \\
    \hline
    3 Centers, $\Sigma_1$ & 0.9912 & 0.9948 & 0.9932 & 0.9897 \\
    \hline
    3 Centers, $\Sigma_2$ & 0.8852 & 0.9181 & 0.8906 & 0.8676 \\
    \hline
    3 Centers, $\Sigma_3$ & 0.6848 & 0.7104 & 0.6714 & 0.6391 \\
    \hline
    4 Centers, $\Sigma_1$ & 0.9855 & 0.9923 & 0.9913 & 0.9848 \\
    \hline
    4 Centers, $\Sigma_2$ & 0.7025 & 0.7844 & 0.7824 & 0.6974 \\
    \hline
    4 Centers, $\Sigma_3$ & 0.3607 & 0.4282 & 0.4311 & 0.3491 \\
    \hline
    Random 2 Centers & -0.0047 & -0.0043 & -0.0042 & -0.0094 \\
    \hline
    Random 3 Centers & -0.0055 & -0.0062 & -0.0076 & -0.0167 \\
    \hline
    Random 4 Centers & 0.0005 & 0.0011 & 0.0008 & -0.0188 \\
    \hline
    \end{tabular}
    \label{table:fadjst}
    \end{table}

\subsection{Comparing two fuzzy partitions}
\label{fuzzyfuzzy}

\noindent For the second simulation study, we generated $7$ data sets with $\mathbf{C}=2,3,\ldots,8$ cluster centers by incrementally merging eight different bi-variate normal distributions with mean vectors\\

\begin{center}
\begin{tabular}{l}
$\boldsymbol{\mu}_1 = [-2,-2];$ \\
$\boldsymbol{\mu}_2 = [2,2];$ \\
$\boldsymbol{\mu}_3 = [0,0];$ \\
$\boldsymbol{\mu}_4 = [-2,2];$ \\
$\boldsymbol{\mu}_5 = [2,-2];$ \\
$\boldsymbol{\mu}_6 = [-4,4];$ \\
$\boldsymbol{\mu}_7 = [4,-4];$ \\
$\boldsymbol{\mu}_8 = [9, 9];$ \\
\end{tabular}
\end{center}

\noindent and covariance matrix equal to $\textrm{I}\times \boldsymbol{\alpha}'$, where $\boldsymbol{\alpha}$ is a vector made of two draws from a uniform distribution in $(0.1,1)$. The structure of the data sets is presented in Table \ref{tab0}. The sample size was set, this time, to 120.

\begin{table}[h]
\centering
\caption{Second simulation study: data sets structure}
\begin{tabular}{|l|l|}
\hline
\multicolumn{1}{|c|}{Data set number} & \multicolumn{1}{c|}{Centers} \\
\hline
\multicolumn{1}{|c|}{1} & $\boldsymbol{\mu}_1, \textrm{ }\boldsymbol{\mu}_2$ \\
\hline
\multicolumn{1}{|c|}{2} & Add $\boldsymbol{\mu}_3$ to data set 1\\
\hline
\multicolumn{1}{|c|}{3} & Add $\boldsymbol{\mu}_4$ to data set 2 \\
\hline
\multicolumn{1}{|c|}{4} & Add $\boldsymbol{\mu}_5$ to data set 3\\
\hline
\multicolumn{1}{|c|}{5} & Add $\boldsymbol{\mu}_6$ to data set 4\\
\hline
\multicolumn{1}{|c|}{6} & Add $\boldsymbol{\mu}_7$ to data set 5\\
\hline
\multicolumn{1}{|c|}{7} & Add $\boldsymbol{\mu}_8$ to data set 6\\
\hline
\end{tabular}
\label{tab0}
\end{table}

\noindent For each data set we ran the fuzzy \textit{C}-means algorithm by setting $C=2,3,\ldots,8$.
Then we computed \citeauthor{huller} NDC and the ACI between the returned fuzzy partitions for each data set and for $C=2,3,\ldots,8$.
We decided to not show in this case the other variants for Rand index and Adjusted Rand index by \cite{brouwer,campello} and \cite{anderson} since, even if they can be applied to compare two fuzzy partitions, they lead to inconclusive measures (i.e.\ these indexes do not satisfy the reflexivity condition. For further details, we refer to \cite{campello, anderson, huller}.

    \begin{table}[ht]
    \centering
    \caption{Comparing two fuzzy partitions: normalized degree of concordance}
    \begin{tabular}{|c|c|c|c|c|c|c|c|}
    \hline
     & C2 & C=3 & C=4 & C=5 & C=6 & C=7 & C=8 \\
    \hline
    Data set 1 & 1.0000 & 0.9012 & 0.8423 & 0.7895 & 0.7607 & 0.7558 & 0.7236 \\
    \hline
    Data set 2 & 0.8125 & 1.0000 & 0.8945 & 0.8853 & 0.8610 & 0.8252 & 0.8182 \\
    \hline
    Data set 3 & 0.7027 & 0.8681 & 1.0000 & 0.9304 & 0.9083 & 0.8925 & 0.8687 \\
    \hline
    Data set 4 & 0.7024 & 0.8346 & 0.9201 & 1.0000 & 0.9668 & 0.9291 & 0.9157 \\
    \hline
    Data set 5 & 0.6850 & 0.8231 & 0.8820 & 0.9275 & 1.0000 & 0.9511 & 0.9286 \\
    \hline
    Data set 6 & 0.6173 & 0.7582 & 0.8438 & 0.8944 & 0.9513 & 1.0000 & 0.9728 \\
    \hline
    Data set 7 & 0.4540 & 0.6973 & 0.8084 & 0.8776 & 0.9177 & 0.9551 & 1.0000 \\
    \hline
    \end{tabular}
    \label{table:NDCrand}
    \end{table}
		
    \begin{table}[ht]
    \centering
		 \caption{Comparing two fuzzy partitions: Adjusted Concordance index}
    \begin{tabular}{|c|c|c|c|c|c|c|c|}
    \hline
    & C=2 & C=3 & C=4 & C=5 & C=6 & C=7 & C=8 \\
    \hline
    Data set 1 & 1.0000 & 0.7895 & 0.6606 & 0.5484 & 0.4879 & 0.4768 & 0.4093 \\
    \hline
    Data set 2 & 0.5550 & 1.0000 & 0.7196 & 0.6889 & 0.6188 & 0.5092 & 0.4870 \\
    \hline
    Data set 3 & 0.2743 & 0.6221 & 1.0000 & 0.7774 & 0.7000 & 0.6416 & 0.5539 \\
    \hline
    Data set 4 & 0.3319 & 0.5603 & 0.7685 & 1.0000 & 0.8964 & 0.7652 & 0.7188 \\
    \hline
    Data set 5 & 0.2624 & 0.4906 & 0.6230 & 0.7530 & 1.0000 & 0.8235 & 0.7318 \\
    \hline
    Data set 6 & 0.1755 & 0.3461 & 0.5082 & 0.6399 & 0.8201 & 1.0000 & 0.8927 \\
    \hline
    Data set 7 & 0.1002 & 0.2751 & 0.4457 & 0.5863 & 0.6985 & 0.8254 & 1.0000 \\
    \hline
    \end{tabular}
    \label{table:sim2_ACI}
    \end{table}

\noindent Both Tables \ref{table:NDCrand} and \ref{table:sim2_ACI} show that NDC and ACI are equal to 1 when comparing the same partitions. In both Tables we note that the value of the indexes is larger when the number of estimated partitions is close to the true generated partitions. It can be noted that both indexes tend to return slightly larger values when estimating a number of partitions higher than number of true partitions. But the magnitude of the ACI is more realistic than the one of the NDC. For example, in the first row of Table \ref{table:NDCrand}, the value of the NDC is equal to $0.7236$ when we compare the fuzzy partitions of a data set with two centers with the solutions with C=8 estimated centers, while in the same situation (first row of Table \ref{table:sim2_ACI}) ACI=0.4093.
The same conclusion can be reached when we compare the last row of these Tables (NDC=0.4540, ACI=0.1002).
It is worth noting that these differences in magnitude are due to the fact that while the NDC still considers the partitions quite similar to each other and, for this reason, not so far away, the ACI, by estimating the expected value of the index to take into account the model of randomness, informs that part of the similarity detected by NDC is due to chance.
It is worth highlighting that this cannot be stated for the extensions of the Adjusted Rand index proposed by the other authors since they used the standard formulation of the ARI by modifying the four cardinalities, but this does not guarantee that the expected values of their fuzzy Rand indexes is correctly identified.

\subsection{Comparing estimated and ``true'' fuzzy partitions}
\label{PDclust}
\noindent To explain what we mean for "true" fuzzy partition we introduce the Probabilistic-Distance (PD) clustering \citep{pdc}.
The PD clustering allows for a probabilistic allocation of cases to classes or clusters. It is a form of fuzzy clustering that is independent on the specification of fuzzifiers. It is based on the principle that probability and distance are inversely related
\begin{equation}
\label{relation}
p_k(\mathbf{x}) d_k(\mathbf{x}) = constant, \mbox{ depending on } \mathbf{x},
\end{equation}
\noindent in which $d$ is a distance measure between the $j$-th individual and the $k$-th cluster center and $p_k(\mathbf{x})$ denotes the probability of the $j$-th individual to belong to the $k$-th cluster, for $k=1, \ldots, K$.

\noindent Equation \ref{relation} allows to define the membership probabilities as
\begin{equation}
\label{pd}
p_k(\mathbf{x})=\frac{\prod_{j\neq k}{d(\mathbf{x})}}{\sum_{t=1}^K{\prod_{j\neq t}{d(\mathbf{x})}}}, \; { k = 1, \cdots, K }.
\end{equation}

\noindent Using Equation \ref{pd}, we are able to determine for a real or a simulated data set the ``true'' fuzzy partitions, provided that we know a priori the composition of the clusters, and then to compute the indexes between the estimated and ``true'' fuzzy partitions. We should note that we removed all categorical variables from these data and we used always the same setting based on Euclidean distance.\\
\noindent As a first experiment we ran the algorithms on the same data sets used before. Even if we decided to not use in the previous experiment the fuzzy extensions of the indexes proposed by \cite{brouwer, campello, anderson} for comparing fuzzy partitions, in this case we also included these indexes. By looking at both Tables \ref{trf1} and \ref{trf3}, it seems that all these indexes have a similar behavior. Note that, while it is expected that the lower the variance the lower the magnitude of the indexes, it seems that there is a considerable difference among the indexes for different levels of the covariance matrix. Moreover, as also pointed out by \cite{anderson}, these indexes are not comparable. 
\begin{table}[h!]
  \centering
  \caption{Comparing fuzzy partitions: comparison of the estimated membership probabilities and the true fuzzy partition.}
\label{trf1}%
    \begin{tabular}{|r|r|r|r|r|r|r|}
    \hline
          & \multicolumn{3}{c|}{NDC} & \multicolumn{3}{c|}{ACI} \\
    \hline
    data set & $\Sigma_1$ & $\Sigma_2$ & $\Sigma_3$ & $\Sigma_1$ & $\Sigma_2$ & $\Sigma_3$ \\
    \hline
    \multicolumn{1}{|c|}{2 centers} & 0.9988 & 0.9922 & 0.9808 & 0.9975 & 0.9810 & 0.9413 \\
    \hline
    \multicolumn{1}{|c|}{3 centers} & 0.9963 & 0.9834 & 0.9812 & 0.9908 & 0.9524 & 0.9405 \\
    \hline
    \multicolumn{1}{|c|}{4 centers} & 0.8557 & 0.9639 & 0.8915 & 0.6201 & 0.8639 & 0.5038 \\
    \hline
    \multicolumn{1}{|c|}{} & \multicolumn{3}{c|}{Brouwer Rand} & \multicolumn{3}{c|}{Brouwer Adjusetd Rand} \\
    \hline
    \multicolumn{1}{|c|}{2 centers} & 0.959507 & 0.834307 & 0.760977 & 0.918929 & 0.654738 & 0.429033 \\
    \hline
    \multicolumn{1}{|c|}{3 centers} & 0.913283 & 0.762435 & 0.697006 & 0.814727 & 0.524424 & 0.37679 \\
    \hline
    \multicolumn{1}{|c|}{4 centers} & 0.712589 & 0.669445 & 0.652062 & 0.423158 & 0.337953 & 0.191748 \\
    \hline
    \multicolumn{1}{|c|}{} & \multicolumn{3}{c|}{Campello Rand} & \multicolumn{3}{c|}{Campello Adjusted Rand} \\
    \hline
    \multicolumn{1}{|c|}{2 centers} & 0.950726 & 0.804091 & 0.686992 & 0.901445 & 0.608165 & 0.373974 \\
    \hline
    \multicolumn{1}{|c|}{3 centers} & 0.91354 & 0.757514 & 0.665499 & 0.811207 & 0.499784 & 0.325355 \\
    \hline
    \multicolumn{1}{|c|}{4 centers} & 0.73316 & 0.701056 & 0.584516 & 0.424566 & 0.370523 & 0.156312 \\
    \hline
    \multicolumn{1}{|c|}{} & \multicolumn{3}{c|}{Anderson Rand} & \multicolumn{3}{c|}{Anderson Adjusted Rand} \\
    \hline
    \multicolumn{1}{|c|}{2 centers} & 0.922612 & 0.718153 & 0.592507 & 0.845209 & 0.436248 & 0.184937 \\
    \hline
    \multicolumn{1}{|c|}{3centers} & 0.854373 & 0.69217 & 0.63614 & 0.668937 & 0.307852 & 0.203322 \\
    \hline
    \multicolumn{1}{|c|}{4 centers} & 0.732508 & 0.667434 & 0.632849 & 0.289444 & 0.095834 & 0.002191 \\
    \hline
    \end{tabular}%
  
\end{table}%

\begin{table}[h!]
  \centering
  \caption{Comparing fuzzy partitions: both partitions are the true fuzzy partitions.}
\label{trf2}%
    \begin{tabular}{|c|r|r|r|r|r|r|}
    \hline
          & \multicolumn{3}{c|}{NDC} & \multicolumn{3}{c|}{ACI} \\
    \hline
    data set & $\Sigma_1$ & $\Sigma_2$ & $\Sigma_3$ & $\Sigma_1$ & $\Sigma_2$ & $\Sigma_3$ \\
    \hline
    2 centers & 1.0000 & 1.0000 & 1.0000 & 1.0000 & 1.0000 & 1.0000 \\
    \hline
    3 centers & 1.0000 & 1.0000 & 1.0000 & 1.0000 & 1.0000 & 1.0000 \\
    \hline
    4 centers & 1.0000 & 1.0000 & 1.0000 & 1.0000 & 1.0000 & 1.0000 \\
    \hline
          & \multicolumn{3}{c|}{Brouwer Rand} & \multicolumn{3}{c|}{Brouwer Adjusetd Rand} \\
    \hline
    2 centers & 0.9595 & 0.8344 & 0.7610 & 0.9189 & 0.6550 & 0.4287 \\
    \hline
    3 centers & 0.9132 & 0.7619 & 0.6971 & 0.8146 & 0.5233 & 0.3768 \\
    \hline
    4 centers & 0.8947 & 0.6681 & 0.6623 & 0.7493 & 0.3349 & 0.2127 \\
    \hline
          & \multicolumn{3}{c|}{Campello Rand} & \multicolumn{3}{c|}{Campello Adjusted Rand} \\
    \hline
    2 centers & 0.9508 & 0.8055 & 0.6913 & 0.9016 & 0.6109 & 0.3826 \\
    \hline
    3 centers & 0.9138 & 0.7597 & 0.6692 & 0.8119 & 0.5044 & 0.3329 \\
    \hline
    4 centers & 0.9112 & 0.7056 & 0.6124 & 0.7830 & 0.3801 & 0.2112 \\
    \hline
          & \multicolumn{3}{c|}{Anderson Rand} & \multicolumn{3}{c|}{Anderson Adjusted Rand} \\
    \hline
    2 centers & 0.9225 & 0.7182 & 0.5919 & 0.8451 & 0.4363 & 0.1837 \\
    \hline
    3 centers & 0.8542 & 0.6912 & 0.6357 & 0.6685 & 0.3057 & 0.2024 \\
    \hline
    4 centers & 0.8419 & 0.6658 & 0.6331 & 0.5696 & 0.0913 & 0.0018 \\
    \hline
    \end{tabular}%
  
\end{table}%

Both Tables \ref{trf2} and  \ref{trf4} show the indexes computed comparing the true fuzzy partition with itself. In this case, the reflexivity property of the \citeauthor{huller}'s NDC is emphasized. We think now it is straightforward that the adjusted versions of the fuzzy Rand-like indexes do not make sense if these are computed using Formula \ref{ari}. In some cases the Adjusted Rand-like index comparing the same partition is negative and it should be set equal to zero (as in the case of the \citeauthor{anderson}'s index in Table \ref{trf4}).

\begin{table}[h!]
  \centering
	\scriptsize
  \caption{Comparing fuzzy partitions: comparison of the estimated membership probabilities and the true fuzzy partition.}
  \label{trf3}%
    \begin{tabular}{|r|r|r|r|r|r|r|r|r|}
    \hline
    data set & \multicolumn{1}{c|}{NDC} & \multicolumn{1}{c|}{ACI} & \multicolumn{1}{c|}{Brouwer} & \multicolumn{1}{c|}{Brouwer} & \multicolumn{1}{c|}{Campello} & \multicolumn{1}{c|}{Campello} & \multicolumn{1}{c|}{Anderson} & \multicolumn{1}{c|}{Anderson} \\
          & \multicolumn{1}{c|}{} & \multicolumn{1}{c|}{} & \multicolumn{1}{c|}{ Rand} & \multicolumn{1}{c|}{Adj. Rand} & \multicolumn{1}{c|}{Rand} & \multicolumn{1}{c|}{Adj. Rand} & \multicolumn{1}{c|}{Rabd} & \multicolumn{1}{c|}{Adj. Rand} \\
    \hline
    Random1 & 0.8125 & -0.0055 & 0.8789 & -0.0017 & 0.4999 & -0.0002 & 0.4949 & -0.0102 \\
    \hline
    Random2 & 0.7836 & 0.0478 & 0.7867 & 0.0130 & 0.5106 & 0.0206 & 0.5514 & -0.0195 \\
    \hline
    Random3 & 0.8250 & 0.1242 & 0.7685 & 0.0450 & 0.5152 & 0.0283 & 0.6218 & -0.0290 \\
    \hline
    \end{tabular}%

\end{table}%

\begin{table}[h!]
  \centering
	\scriptsize
  \caption{Comparing fuzzy partitions: both partitions are the true fuzzy partitions.}
  \label{trf4}%
    \begin{tabular}{|r|r|r|r|r|r|r|r|r|}
    \hline
    data set & \multicolumn{1}{c|}{NDC} & \multicolumn{1}{c|}{ACI} & \multicolumn{1}{c|}{Brouwer} & \multicolumn{1}{c|}{Brouwer} & \multicolumn{1}{c|}{Campello} & \multicolumn{1}{c|}{Campello} & \multicolumn{1}{c|}{Anderson} & \multicolumn{1}{c|}{Anderson} \\
          & \multicolumn{1}{c|}{} & \multicolumn{1}{c|}{} & \multicolumn{1}{c|}{ Rand} & \multicolumn{1}{c|}{Adj. Rand} & \multicolumn{1}{c|}{Rand} & \multicolumn{1}{c|}{Adj. Rand} & \multicolumn{1}{c|}{Rabd} & \multicolumn{1}{c|}{Adj. Rand} \\
    \hline
    Random1 & 1.0000 & 1.0000 & 0.9886 & 0.0141 & 0.5086 & 0.0173 & 0.4950 & -0.0102 \\
    \hline
    Random2 & 1.0000 & 1.0000 & 0.9493 & 0.0599 & 0.5147 & 0.0293 & 0.5512 & -0.0202 \\
    \hline
    Random3 & 1.0000 & 1.0000 & 0.8661 & 0.1352 & 0.5271 & 0.0532 & 0.6216 & -0.0294 \\
    \hline
    \end{tabular}%

\end{table}%

As a second experiment we used real data sets taken from the UCI repository for machine learning. 
Results are summarized in Table \ref{tab5}.

\begin{table}[htbp]
  \centering
	\small
	\caption{Comparison between estimated and true probabilistic partitions. Extended Rand indexes, on the left side, and extended adjusted Rand indexes on the right side. For each side of the table, the first column represents the index computed between the true probabilistic partition and itself, while the second column represents the index computed between the true probabilistic partition and the one returned by the PD-clustering algorithm by setting the number of clusters equal to the known number of clusters, indicated in brackets}
		\label{tab5}
    \begin{tabular}{|c|r|r|r|r|r|r|}
    \hline
          & \multicolumn{1}{c|}{} & \multicolumn{1}{c|}{True} & \multicolumn{1}{c|}{C} & \multicolumn{1}{c|}{} & \multicolumn{1}{c|}{True} & \multicolumn{1}{c|}{C} \\
    Data sets & \multicolumn{1}{c|}{Rand } & \multicolumn{1}{c|}{fuzzy} & \multicolumn{1}{c|}{clusters} & \multicolumn{1}{c|}{Adjusted} & \multicolumn{1}{c|}{fuzzy} & \multicolumn{1}{c|}{clusters} \\
          & \multicolumn{1}{c|}{extensions} & \multicolumn{1}{c|}{partition} & \multicolumn{1}{c|}{solution} & \multicolumn{1}{c|}{extensions} & \multicolumn{1}{c|}{partition} & \multicolumn{1}{c|}{solution} \\
    \hline
    \multicolumn{1}{|c|}{} & \multicolumn{1}{l|}{Anderson} & 0.6227 & 0.6261 & \multicolumn{1}{l|}{Anderson} & 0.0223 & 0.0272 \\
    \multicolumn{1}{|c|}{Vehicle} & \multicolumn{1}{l|}{Campello} & 0.5699 & 0.5325 & \multicolumn{1}{l|}{Campello} & 0.1397 & 0.0656 \\
    \multicolumn{1}{|c|}{(C=4)} & \multicolumn{1}{l|}{Brouwer} & 0.7670 & 0.6526 & \multicolumn{1}{l|}{Brouwer} & 0.2159 & 0.1639 \\
    \multicolumn{1}{|c|}{} & \multicolumn{1}{l|}{NDC} & 1.0000 & 0.8085 & \multicolumn{1}{l|}{ACI} & 1.0000 & 0.3008 \\
    \hline
    \multicolumn{1}{|c|}{} & \multicolumn{1}{l|}{Anderson } & 0.4976 & 0.4976 & \multicolumn{1}{l|}{Anderson } & -0.0048 & -0.0049 \\
    \multicolumn{1}{|c|}{Sonar} & \multicolumn{1}{l|}{Campello } & 0.5072 & 0.5000 & \multicolumn{1}{l|}{Campello } & 0.0144 & 0.0000 \\
    \multicolumn{1}{|c|}{(C=2)} & \multicolumn{1}{l|}{Brouwer } & 0.9924 & 0.9962 & \multicolumn{1}{l|}{Brouwer } & 0.0070 & 0.0000 \\
    \multicolumn{1}{|c|}{} & \multicolumn{1}{l|}{NDC} & 1.0000 & 0.9647 & \multicolumn{1}{l|}{ACI} & 1.0000 & -0.0001 \\
    \hline
    \multicolumn{1}{|c|}{} & \multicolumn{1}{l|}{Anderson } & 0.5003 & 0.5025 & \multicolumn{1}{l|}{Anderson } & 0.0007 & 0.0050 \\
    \multicolumn{1}{|c|}{Pima} & \multicolumn{1}{l|}{Campello } & 0.5333 & 0.5265 & \multicolumn{1}{l|}{Campello } & 0.0665 & 0.0528 \\
    \multicolumn{1}{|c|}{(C=2)} & \multicolumn{1}{l|}{Brouwer } & 0.9365 & 0.8039 & \multicolumn{1}{l|}{Brouwer } & 0.0648 & 0.0407 \\
    \multicolumn{1}{|c|}{} & \multicolumn{1}{l|}{NDC} & 1.0000 & 0.8125 & \multicolumn{1}{l|}{ACI} & 1.0000 & 0.1431 \\
    \hline
    \multicolumn{1}{|c|}{} & \multicolumn{1}{l|}{Anderson } & 0.6167 & 0.6160 & \multicolumn{1}{l|}{Anderson } & 0.1321 & 0.1305 \\
    \multicolumn{1}{|c|}{Iris} & \multicolumn{1}{l|}{Campello } & 0.6552 & 0.6477 & \multicolumn{1}{l|}{Campello } & 0.3005 & 0.2860 \\
    \multicolumn{1}{|c|}{(C=3)} & \multicolumn{1}{l|}{Brouwer } & 0.7211 & 0.7229 & \multicolumn{1}{l|}{Brouwer } & 0.4096 & 0.4107 \\
    \multicolumn{1}{|c|}{} & \multicolumn{1}{l|}{NDC} & 1.0000 & 0.9780 & \multicolumn{1}{l|}{ACI} & 1.0000 & 0.9298 \\
    \hline
    \multicolumn{1}{|c|}{} & \multicolumn{1}{l|}{Anderson } & 0.4989 & 0.4997 & \multicolumn{1}{l|}{Anderson } & -0.0022 & -0.0005 \\
    \multicolumn{1}{|c|}{Ionosphere} & \multicolumn{1}{l|}{Campello } & 0.5185 & 0.5168 & \multicolumn{1}{l|}{Campello } & 0.0371 & 0.0336 \\
    \multicolumn{1}{|c|}{(C=2)} & \multicolumn{1}{l|}{Brouwer } & 0.9596 & 0.9028 & \multicolumn{1}{l|}{Brouwer } & 0.0390 & 0.0440 \\
    \multicolumn{1}{|c|}{} & \multicolumn{1}{l|}{NDC} & 1.0000 & 0.8979 & \multicolumn{1}{l|}{ACI} & 1.0000 & 0.2488 \\
    \hline
    \multicolumn{1}{|c|}{} & \multicolumn{1}{l|}{Anderson } & 0.7333 & 0.7317 & \multicolumn{1}{l|}{Anderson } & 0.0103 & 0.0108 \\
    \multicolumn{1}{|c|}{Flags} & \multicolumn{1}{l|}{Campello } & 0.5823 & 0.5573 & \multicolumn{1}{l|}{Campello } & 0.1575 & 0.0938 \\
    \multicolumn{1}{|c|}{(C=8)} & \multicolumn{1}{l|}{Brouwer } & 0.7300 & 0.5301 & \multicolumn{1}{l|}{Brouwer } & 0.3437 & 0.1202 \\
    \multicolumn{1}{|c|}{} & \multicolumn{1}{l|}{NDC} & 1.0000 & 0.6828 & \multicolumn{1}{l|}{ACI} & 1.0000 & 0.1448 \\
    \hline
    \end{tabular}%
\end{table}%

We would like to point out that it is not our intention to use the extensions of the Rand and of the Adjusted Rand index to identify the number of clusters to use since these indexes are external validity measures. As can be noted from Table \ref{tab5} and as already pointed out, all the indexes, apart from NDC and ACI, do not respect the reflexivity property. Furthermore, in some cases, the value of the index estimated when we compare the same partition (first and third column) is lower than the value obtained when comparing the estimated partition with the true partition (e.g.\ in the Vehicle data set the \citeauthor{anderson} approach for both the Rand and the adjusted Rand extensions). Nevertheless, even if it is not correct to compare these indexes (as also stated in\cite{anderson}), it seems that the behavior of the corrections applied to the adjusted extensions with respect to its Rand extensions is quite similar. Of course, the interpretation of all these corrections as a correction for randomness is possible, as already stated, only when we consider our approach with respect to NDC. For instance, in the case of the Sonar data set, NDC is equal to $0.9647$ when comparing the true and the estimated partition and ACI is equal to $-0.0001$. These results could seem inconsistent, but if we take into account the true partition, we can notice that each object has a probability to belong to each cluster that is really close to $0.5$ and the membership probabilities estimated by the PD-clustering algorithm are also close to $0.5$. In sight of this it is obvious that a NDC close to 1 means that these two partitions are really close to each other. On the other hand, probabilities close to $0.5$ are equivalent to a coin flipping experiment and, for this reason, it is not surprising that ACI is close to zero. On the contrary, for the Iris data set, NDC is $0.9780$ and ACI is $0.9298$. These results indicate that the estimated partition is really close to the true one, but, in this case, this similarity is not due to chance.
It is worth stressing again that our goal is not to evaluate clustering algorithm, but only the behaviour of these external validation indexes.

\noindent

\section{Concluding remarks}
\label{final}
\noindent In this paper we proposed the adjusted version of the normalized degree of concordance index (NDC) defined by \citet{huller} for comparing fuzzy partitions, namely the Adjusted Concordance Index (ACI).
This measure is constructed upon a similar reasoning of the well known Adjusted Rand index by \cite{hubert} when applied to compare hard partitions.\\
We derived the proposed index by normalizing the difference between NDC and its expected value obtained by considering a large number of permutation of the similarities considered in the similarity matrices.
Experimental evaluations show that ACI returns more coherent results than the NDC in comparing fuzzy partitions when the aim is to validate clustering solutions, because it takes into account the possible randomness component of the similarity measure. The same approach cannot be applied to the other indexes because them are not reflexive and their maximum value is not known. Moreover, when comparing a fuzzy and a crisp partition, ACI is closely related to the adjusted Rand index for fuzzy partitions defined by \citet{campello, brouwer, anderson}.
It should be noted that ACI can be used in comparing both fuzzy and crisp and only fuzzy partitions.
Furthermore, the approach for dealing with fuzzy partitions introduced by \citet{huller} ensures that ACI shows properties of a proper metric under certain assumptions (i.e.\ Ruspini’s partitions, probabilistic partitions).\\
In addition, using the PD-clustering approach by \cite{pdc}, we were able to build the reference true probabilistic partitions for some real data set to show the ACI potentialities as external validity measure.


\bibliographystyle{spbasic}

\begin{thebibliography}{29}
\providecommand{\natexlab}[1]{#1}
\providecommand{\url}[1]{{#1}}
\providecommand{\urlprefix}{URL }
\expandafter\ifx\csname urlstyle\endcsname\relax
  \providecommand{\doi}[1]{DOI~\discretionary{}{}{}#1}\else
  \providecommand{\doi}{DOI~\discretionary{}{}{}\begingroup
  \urlstyle{rm}\Url}\fi
\providecommand{\eprint}[2][]{\url{#2}}

\bibitem[{Anderberg(2014)}]{anderberg}
Anderberg MR (2014) Cluster Analysis for Applications: Probability and
  Mathematical Statistics: A Series of Monographs and Textbooks, vol~19.
  Academic press

\bibitem[{Anderson et~al(2010)Anderson, Bezdek, Popescu, and Keller}]{anderson}
Anderson DT, Bezdek JC, Popescu M, Keller JM (2010) Comparing fuzzy,
  probabilistic, and possibilistic partitions. Fuzzy Systems, IEEE Transactions
  on 18(5):906--918

\bibitem[{Ben-Israel and Iyigun(2008)}]{pdc}
Ben-Israel A, Iyigun C (2008) Probabilistic d-clustering. Journal of
  Classification 25(1):5--26

\bibitem[{Berkhin(2006)}]{berkhin}
Berkhin P (2006) A survey of clustering data mining techniques. In: Grouping
  multidimensional data, Springer, pp 25--71

\bibitem[{Bezdek et~al(1984)Bezdek, Ehrlich, and Full}]{bezdek}
Bezdek JC, Ehrlich R, Full W (1984) Fcm: The fuzzy c-means clustering
  algorithm. Computers \& Geosciences 10(2):191--203

\bibitem[{B{\"o}ck(1974)}]{bock}
B{\"o}ck HH (1974) Automatische Klassifikation: theoret. u. prakt. Methoden z.
  Gruppierung u. Strukturierung von Daten (Cluster-Analyse), vol~24.
  Vandenhoeck \& Ruprecht

\bibitem[{Brouwer(2009)}]{brouwer}
Brouwer RK (2009) Extending the rand, adjusted rand and jaccard indices to
  fuzzy partitions. Journal of Intelligent Information Systems 32(3):213--235

\bibitem[{Campello(2007)}]{campello}
Campello RJ (2007) A fuzzy extension of the rand index and other related
  indexes for clustering and classification assessment. Pattern Recognition
  Letters 28(7):833--841

\bibitem[{Dice(1945)}]{dice}
Dice LR (1945) Measures of the amount of ecologic association between species.
  Ecology 26(3):297--302

\bibitem[{Downton and Brennan(1980)}]{downton}
Downton M, Brennan T (1980) Comparing classifications: an evaluation of several
  coefficients of partition agreement. Class Soc Bull 4(4):53--54

\bibitem[{Duran and Odell(2013)}]{duran}
Duran BS, Odell PL (2013) Cluster analysis: a survey, vol 100. Springer Science
  \& Business Media

\bibitem[{Fasulo(1999)}]{fasulo}
Fasulo D (1999) An analysis of recent work on clustering algorithms. Department
  of Computer Science \& Engineering, University of Washington

\bibitem[{Fowlkes and Mallows(1983)}]{fowlkes}
Fowlkes EB, Mallows CL (1983) A method for comparing two hierarchical
  clusterings. Journal of the American statistical association 78(383):553--569

\bibitem[{Frigui et~al(2007)Frigui, Hwang, and Rhee}]{frigui}
Frigui H, Hwang C, Rhee FCH (2007) Clustering and aggregation of relational
  data with applications to image database categorization. Pattern Recognition
  40(11):3053--3068

\bibitem[{Hartigan(1975)}]{hartigan}
Hartigan JA (1975) Clustering algorithms. John Wiley \& Sons, Inc.

\bibitem[{H{\"o}ppner et~al(1999)H{\"o}ppner, Klawonn, Kruse, and
  Runkler}]{hoppner}
H{\"o}ppner F, Klawonn F, Kruse R, Runkler T (1999) Fuzzy cluster analysis:
  methods for classification, data analysis and image recognition. 1999.
  Baffins Lan, Chichester, West Sussex, PO19 1UD, England: Lohn Wiley \& Sons
  Ltd

\bibitem[{Hubert and Arabie(1985)}]{hubert}
Hubert L, Arabie P (1985) Comparing partitions. Journal of classification
  2(1):193--218

\bibitem[{H\"ullermeier et~al(2012)H\"ullermeier, Rifqi, Henzgen, and
  Senge}]{huller}
H\"ullermeier E, Rifqi M, Henzgen S, Senge R (2012) Comparing fuzzy partitions:
  A generalization of the rand index and related measures. Fuzzy Systems, IEEE
  Transactions on 20(3):546--556

\bibitem[{Jain and Dubes(1988)}]{jain88}
Jain AK, Dubes RC (1988) Algorithms for clustering data. Prentice-Hall, Inc.

\bibitem[{Jain et~al(1999)Jain, Murty, and Flynn}]{jain99}
Jain AK, Murty MN, Flynn PJ (1999) Data clustering: a review. ACM computing
  surveys (CSUR) 31(3):264--323

\bibitem[{Kaufman and Rousseeuw(2009)}]{kaufman}
Kaufman L, Rousseeuw PJ (2009) Finding groups in data: an introduction to
  cluster analysis, vol 344. John Wiley \& Sons

\bibitem[{Klement et~al(2013)Klement, Mesiar, and Pap}]{klement}
Klement EP, Mesiar R, Pap E (2013) Triangular norms, vol~8. Springer Science \&
  Business Media

\bibitem[{Meil{\u{a}}(2007)}]{meilua}
Meil{\u{a}} M (2007) Comparing clusterings - an information based distance.
  Journal of multivariate analysis 98(5):873--895

\bibitem[{Mirkin(1998)}]{mirkin}
Mirkin B (1998) Mathematical classification and clustering: From how to what
  and why. Springer

\bibitem[{Morey and Agresti(1984)}]{morey}
Morey LC, Agresti A (1984) The measurement of classification agreement: an
  adjustment to the rand statistic for chance agreement. Educational and
  Psychological Measurement 44(1):33--37

\bibitem[{Rand(1971)}]{rand}
Rand WM (1971) Objective criteria for the evaluation of clustering methods.
  Journal of the American Statistical association 66(336):846--850

\bibitem[{Ruspini(1970)}]{ruspini}
Ruspini EH (1970) Numerical methods for fuzzy clustering. Information Sciences
  2(3):319--350

\bibitem[{Spath(1980)}]{spath}
Spath H (1980) Cluster analysis algorithms for data reduction and
  classification of objects. Ellis Horwood, Ltd. Chichester, England

\bibitem[{Wagner and Wagner(2007)}]{wagner}
Wagner S, Wagner D (2007) Comparing clusterings: an overview. Universit{\"a}t
  Karlsruhe, Fakult{\"a}t f{\"u}r Informatik Karlsruhe

\end{thebibliography}

\end{document}